\documentclass{ecai} 

\usepackage{latexsym}
\usepackage{amssymb}
\usepackage{amsmath}
\usepackage{amsthm}
\usepackage{mathtools}
\usepackage{mathrsfs}
\usepackage{booktabs}
\usepackage{enumitem}
\usepackage{graphicx}
\usepackage{color}
\usepackage{physics}
\usepackage{tikz}
\usepackage{dsfont}
\usepackage{pgfplots}
\usetikzlibrary{intersections}

\newtheorem{theorem}{Theorem}
\newtheorem{lemma}[theorem]{Lemma}
\newtheorem{corollary}[theorem]{Corollary}
\newtheorem{proposition}[theorem]{Proposition}

\newtheorem{definition}{Definition}
\newtheorem{notation}[definition]{Notation}

\newtheorem{example}[theorem]{Example}

\newcommand{\BibTeX}{B\kern-.05em{\sc i\kern-.025em b}\kern-.08em\TeX}
\newcommand{\ov}[2]{\pi^{#1}_{#2}}

\newcount\Comments  
\newcommand{\kibitz}[2]{\ifnum\Comments=1{\color{#1}{#2}}\fi}
\newcommand{\rmr}[1]{\kibitz{red}{[RESHEF:#1]}}
\newcommand{\jwr}[1]{\kibitz{blue}{[JONNY:#1]}}

\def\vx{\vec{x}}
\def\tr{T}
\def\ar{F}
\def\R{\mathbb{R}}

\def\el{\mathbf{EL}}
\def\wl{\mathbf{WL}}

\begin{document}


\begin{frontmatter}


\paperid{123} 


\title{Distribution Aggregation via Continuous Thiele's Rules}


\author
{\fnms{Jonathan}~\snm{Wagner}\thanks{Corresponding Author. Email: sjwagner@campus.technion.ac.il}}
\author
{\fnms{Reshef}~\snm{Meir}}

\address
{Technion---Israel Institute of Technology }


\begin{abstract}
We introduce the class of \textit{Continuous Thiele's Rules} that generalize the familiar \textbf{Thiele's rules} \cite{janson2018phragmens} of multi-winner voting to distribution aggregation problems. Each rule in that class maximizes $\sum_if(\ov{i}{})$ where $\ov{i}{}$ is an agent $i$'s satisfaction and $f$ could be any twice differentiable, increasing and concave real function. Based on a single quantity we call the \textit{'Inequality Aversion'} of $f$ (elsewhere known as "Relative Risk Aversion"), we derive bounds on the Egalitarian loss, welfare loss and the approximation of \textit{Average Fair Share}, leading to a quantifiable, continuous presentation of their inevitable trade-offs. In particular, we show that the Nash Product Rule satisfies\textit{ Average Fair Share} in our setting.  
\end{abstract}

\end{frontmatter}


\section{Introduction}

\rmr{note there is some recent literature on fairness-welfare tradeoff in discrete setting, e.g. Lackner and Skowron MWV paper and Roy's followup paper on PB}

In distribution aggregation problems \cite{elkind2023settling} we study the problem of reaching a collective decision that concerns the division of some continuous public resource into different channels, e.g. the public budget allocated between various objectives (or 'projects') \cite{freeman2019truthful}, time or land allocated between different activities, etc. The important defining feature is that we allocate any \textit{public} resource, meaning that all alternatives serve, in principle, everyone, albeit to variable extents according to personal taste. That is in contrast to different branches of Social Choice, e.g. Cake cutting \cite{procaccia2016cake} or Fair Division \cite{moulin2019fair} where we allocate resources among agents which enjoy them individually as private goods. The collective decision is reached via some voting \textit{rule} or 'mechanism' that inputs everyone's preference regarding the different possible outcomes and outputs an outcome that would be implemented.
\subsection{Preference Modeling}
Accordingly, the preference of generic agents in distribution aggregation concerns the distribution as a whole rather than some personal endowment. In general, the full structure of such preference might be complex. Realistically, however, an aggregation rule that requires deep inquiry into each voter's personal preference seems too bothering and not very likely. Thus, as in other areas of Social Choice, \textit{"single-peaked"} preferences where an agent's optimal outcome entails the complete description of her preferences over the decision space, are many times favored as a reasonable compromise between the accuracy and applicability \cite{moulin1980strategy,brandt2024optimal}. Moreover, such models typically perform better in terms of strategy-proofness. Specifically, $\ell_p$-norm preference where an agent (dis)satisfaction is expressed as the distance $\norm{x^i-x}_p$ between her preferred distribution $x^i$ to the one implemented $x$, and especially $\ell_1$ is probably the most prevailing choice \cite{elkind2023settling,goel2019knapsack,freeman2019truthful}. In our work we adopt the $\ell_1$ model, albeit adhering to its equivalent, less common representation that was introduced in \cite{goel2019knapsack} of  \textit{overlap} utilities. That is, an  agent $i$ enjoys $|x^i \cap x| = \sum_j \min(x^i_j,x_j)$ in allocation $x$, where $j$ covers the set of alternatives. This formulation not only lets us more conveniently relate to satisfaction rather than dissatisfaction, but also mirrors the canonical notion in the discrete multi-winner voting setting of $|A_i \cap W|$, $A_i$ being the set of alternatives approved by agent $i$ and $W$ the winning set. 
\smallskip 


\subsection{Axiomatic Demands And Their Incompatibilities}
A well known fact that dates back as far as to Arrow's Theorem \cite{arrow1950difficulty} is that good axiomatic properties of social choice rules are often contradictory. 
For instance, the utilitarian welfare maximizing allocation $x \in \arg \max \sum_i |x^i \cap x|$ is known to also promise strategyproofness \cite{goel2019knapsack,freeman2019truthful}, while the egalitarian maxmin allocation $x=\max_y \min_i |x^i \cap y|$ can be advocated as the more fair solution and in particular guarantees a $\frac{1}{n}$ satisfaction the least for every agent. The Nash Product Rule $x=\max_y\sum_i \ln(|x^i \cap x|)$ is somewhat a middle ground between the two, that, as we later show, satisfies \textit{Average Justified Representation} (\textbf{AFS}) that cares for the well being of subgroups rather than just individuals. Each of these three admits a certain demand that captures a different angle on social justice. Not surprisingly, they are generically incompatible with each other. In particular, \cite{brandt2024optimal} shows that even the weakest Justified Representation demand of \textit{'proportionality'} is incompatible with welfare maximization and strategyproofness.

\subsubsection{Contribution}
In this work, we offer non-axiomatic approach to deal with these impossibilities, via studying the class   
of \textit{Continuous Theile's Rules} that maximize $\sum_if(|x^i \cap x|)$ where $f$ could be any concave real function.  

The original \textbf{\textit{Thiele's Voting Rules}} \cite{janson2018phragmens} are a well known class of multi-winner voting rules, each characterized by a function $f$, that choose the winning set $W$ to maximize $\sum_i f\Big(|A_i \cap W|\Big)$. In affinity to the discussion above, this class includes some of the most practiced and well studied multi-winner voting rules, e.g. the welfare maximizing $k$-approval with $f=id$ , the egalitarian Chamberlin-Courant with $f=\mathds{1}_{|A_i \cap W| \geq 1}$ , and PAV where $f\Big(|A_i \cap W|\Big)=1+\frac{1}{2}+\frac{1}{3}+\dots+\frac{1}{|A_i \cap W|}$ that satisfies \textit{Extended Justified Representation} \cite{aziz2017justified}. When applied to our continuous setting, much resemblance is found between corresponding rules and the properties they guarantee. The advantage in moving to a continuous space is that we can consider a continuum of rules that lie 'in between' those that properly satisfy different axioms, and characterize the trade-off each of them offers in terms of approximating either axiom. We will show how a rule's approximation guarantees depend on the \textit{Inequality Aversion}  of $f$, $IAV_f=-\frac{tf'(t)}{f''(t)}$ (elsewhere known as 'Relative Risk Aversion' \cite{dyer1982relative}). In particular, we show that in our setup the Nash product rule satisfies \textbf{AFS} by its full meaning.

\subsection{Related Work}
Compared to its neighbouring branches of Social Choice, the literature on distribution aggregation is not abundant. The most natural application is Participatory Budgeting (PB) that has drawn increased attention in late years, however that literature mainly focuses on the discrete form where each project has a fixed cost \cite{peters2021proportional,aziz2021participatory, talmon2019framework}, which is in compliance with the vast majority of PB instances in reality. In some works, the aggregation rule outputs a distribution while preferences are yet binary or ordinal \cite{aziz2019fair,bogomolnaia2005collective,michorzewski2020price}. As customary in Social Choice, the divisible PB literature treats the trade-off between fairness and welfare-optimality mainly by means of offering different mechanisms that can or cannot satisfy specific axiomatic properties \cite{elkind2023settling,aziz2019fair,freeman2019truthful}. \cite{freeman2019truthful} introduces the class of strategy-proof Moving-Phantom mechanisms within which the $(\ell_1)$ welfare maximizing mechanism is the only pareto-efficient one, while the novel "Independent-Market" mechanism is the only one to satisfy a fairly weak notion of proportional fairness. This incompatibility of efficiency, strategy-proofness and proportionality has lately been proven to be generally inevitable \cite{brandt2024optimal} under $\ell_1$ and $\ell_\infty$ preferences. Remarkably, however, \cite{brandt2024optimal} also shows that under \textit{Minimal Quotient} preferences - where the satisfaction of agent $i$ in $x$ equals $\min_{\scriptscriptstyle \{j: x^i_j >0\}}\frac{x_j}{x^i_j}$ - these typically contradicting objectives can be achieved all at once. 
The most closely related predecessor of our work is found in \cite{michorzewski2020price}, where group-fairness and welfare guarantees of different CTR rules (however named differently) are compared, under preference model slightly different than ours.  Our work expands on it by considering egalitarian fairness two, and characterizing the trade-off by a single parameter. Other
  "flexible" approaches for handling  the inherent trade-off include the relaxation of some of the axioms \cite{aziz2019fair,goel2019knapsack},  or searching for approximation guarantees different rules hold for different axioms \cite{fain2016core,fairstein2022welfare}. Under $\ell_1$ preferences in particular,  in \cite{caragiannis2022truthful} and \cite{freeman2024project} fairness is measured as the $\ell_1$ distance of an allocation to the preferences mean. They introduce optimal rules for minimizing that gap within the Moving-Phantom class \cite{freeman2019truthful}, with quantifiable bounds. However, to the best of our knowledge, the introduction of a continuum of parameterized rules to our choice, where each offers a concrete approximation guarantees, have not yet been attempted.   


\section{Preliminaries}
Let $[n]$ be a set of agents and $[m]$ be a set of alternatives, where $[k] := \{1, . . . , k\}$ for
each positive integer $k$. Subsets of $[n]$ are denoted by lowercase letters e.g.  $s \subset [n]$, and in many places we abuse the notation to represent also the subset size. For vectors in $\mathbb{R}^m_+:= \{y \in R^m|y_j \geq 0\ \forall j \in [m]\}$, we denote $|y|=\sum_j y_j$ and $y \leq z$ means $y_j \leq z_j \forall j$. For any tuple $y^1, \dots , y^t$, we define the \emph{overlap} $(y^1\cap \cdots \cap y^t)$ by $(y^1\cap \cdots \cap y^t)_j:=\min_{1 \leq q \leq t}y^q_j$. An \emph{allocation} $x \in \Delta^m:=\{y \in \mathbb{R}^m_+ | \sum_{j=1}^m y_j=1\}$ is a distribution of some continuously divisible resource (e.g. time, money)  among the $m$ alternatives, where we normalize the overall budget to $1$. The preferences of every agent $i$ are expressed by a single allocation $x^i \in \Delta^m$ that she would like to be implemented ideally, and $ \vec{x} := (x^1, \dots , x^n) \in (\Delta^m)^n$ is the preferences \emph{profile}. We denote by $\vec{x}_{-i}$ the partial profile consisting of all agents excluding $i$, and $\vec{x}_{-s}$ accordingly for a subset $s \subset [n]$.  The \emph{satisfaction} of agent $i$ in allocation $x$ is $\ov{i}{x}:=|x^i\cap x| = \sum_j \min(x^i_j,x_j)$. Or, just $\ov{i}{}$ when specifying the allocation is unnecessary. \rmr{maybe maintain consistent notation by always using the agent(s) as superscript. also $\pi^i(x)$ makes more sense.}
\subsection{Aggregation Rules and their Properties}
\begin{definition}
 An \emph{aggregation rule} is a function $\ar: (\Delta^m)^n \to \Delta^m$ that inputs the preferences profile $\vec{x}$ and outputs an allocation $x \in \Delta^m$.   
\end{definition}
 We survey here several properties that may or may not be satisfied by an allocation $x$ given a profile $\vec{x}$. We say that an aggregation rule $\ar$ satisfies any of these properties if for all $\vec{x} \in (\Delta^m)^n$, $\ar(\vec{x})$ satisfies the corresponding property with respect to $\vec{x}$. \rmr{Add references to all existing concepts} \rmr{also better to avoid fancy fonts like $\ar$}
 \begin{itemize}
     \item \emph{Efficiency} (\textbf{EFF}): An allocation $x$ is  \emph{efficient} if no other allocation $y \in \Delta^m$ exists such that $\ov{i}{x} \geq \ov{i}{y} \quad \forall i \in [n]$ with at least one agent for which that inequality is strict.
     \item  \emph{Range Respecting} (\textbf{RR}): An allocation $x$ is \emph{Range Respecting} if $\min_ix^i_j \leq x_j \leq max_i x^i_j\ \forall j \in [m]$. Note that \textbf{EFF} $\implies$ \textbf{RR}. 
     \item \emph{Individual Fair Share} (\textbf{IFS}): An allocation $x$ satisfies \emph{Individual Fair Share} if $\ov{i}{x} \geq \frac{1}{n}\ \forall i \in [n]$. From an agents' individual perspective, that means that $x$ allocates at least $1/n$ of the budget according to her desire.      
     \item \emph{Average Fair Share} (\textbf{AFS}): A subset of agents $s \subset [n]$ is called $\alpha$-\emph{cohesive} for some $0 < \alpha \leq \frac{|s|}{n}$ if $\left|\bigcap_{i \in s} x^i\right| \geq \alpha$.  An allocation $x$ satisfies \emph{Average Fair Share} if for every $\alpha \in (0,1]$, if  $s \subset [n]$ is $\alpha$-cohesive then $\frac{1}{|s|}\sum_{i \in s}\ov{i}{x} \geq \alpha$. \rmr{for every $\alpha$?}
     \item \emph{Core Stability} (\textbf{CS}): An allocation $x$ satisfies \emph{Core Stability} if for every $s \subset [n]$, no $y \in \mathbb{R}^m_+$ exists such that $|y|=\frac{|s|}{n}$ and $\ov{i}{y} \geq \ov{i}{x}\ \forall i \in s$ with strict inequality for at least one member of $s$. \rmr{do you have any core-related results? does Fain positive core result apply in this model?}
     \item \emph{Proportionality} (\textbf{PROP}) \cite{freeman2019truthful}: This much weaker demand requires the fulfilment of all \textit{Fair Share} axioms above, but only for a simple structure of the preference profile (in which they all coincide). \textbf{Single-minded} profiles are profiles where $x^i$ is a unit vector for all $i$, in other words that every agent wishes to allocate the full budget to a single alternative. If $\vx$ is single-minded, then $x$ satisfies \emph{proportionality} if $x=\frac{1}{n}\sum_ix^i$. 
 \end{itemize}
The next two properties refer directly to rules rather than allocations.
\begin{itemize}
    \item \emph{Participation} (\textbf{PAR}): Aggregation rule $\ar$ satisfies (strict) \textit{participation} if 
    \[|\ar(\vec{x})\cap x^i|  \stackrel{(>)}{\geq} |\ar(\vec{x}_{-i})\cap x^i|\quad \forall \vec{x},x^i\]
    meaning that agents should always prefer voting over abstaining. 
    \item \emph{Strategyproofness} (\textbf{SP}): Aggregation rule $\ar$ is \textit{strategyproof } is for all $x^i$ and $\vec{x}_{-i}$,
    \[|\ar(x^i,\vec{x}_{-i})\cap x^i| \geq |\ar(y,\vec{x}_{-i})\cap x^i|\quad \forall y \in \Delta^m\]
     meaning, no agnet $i$ can increase her satisfaction by misreporting some $y \in \Delta^m$ instead of her true preference $x^i$. 
\end{itemize}

\section{Continuous Theile's Rules}
In this paper we study the following class of mechanisms. 
\begin{definition}[Continuous Thiele's Aggregation Rules]
For any increasing, twice differentiable and strictly concave function $f:[0,1] \to \R$, we define the corresponding 
\textbf{Continuous Thiele's Rule} (CTR) $\tr_f$ by 
\[\tr_f(\Vec{x}) \in \arg \max_{x \in \Delta^m} \sum_{i=1}^n f(\ov{i}{x})\]
\end{definition}
\subsection{Optimal allocations}
CTRs can be computed efficiently via convex optimization as $\sum_{i=1}^n f(\ov{i}{x})$ is a concave function of $x$ (note that $\ov{i}{x}=\sum_j\min(x^i_j,x_j)$ is in itself a sum of concave functions). \jwr{Also, all solutions are equivalent in the sense that if $x,y$ are two optima, then $\ov{i}{x}=\ov{i}{y} \forall i$. Important ?}\rmr{not surprising but worth mentioning}. Thanks to that, optimal allocations can be characterized conveniently by first order conditions. We formalize here these constraints that will help us derive much of our later results.
\begin{definition}
    Let $x \in \Delta^m$. For each $j \in [m]$, \rmr{stack the inequality in parentheses}
   \[s^{\uparrow(\downarrow)}_j(x):=\{i \in [n] | x^i_j  > (\geq) x_j\}\]
\end{definition}
That is, $s^{\uparrow}_j(x)$ consists of all agents $i \in [n]$ for which increasing (or reducing, for $s^\downarrow$) $x_j$ while keeping all other $\{x_k\}_{k \neq j}$ fixed will increase (reduce) their satisfaction  $\ov{i}{x}$. The reason we distinguish between the two is that if $x^i_j=x_j$ for some agent $i$, she suffers from a reduction in $x_j$ while not gaining if we increase it. If no such agent exist, the sets coincide and we can just write $s_j$. We also do that for the sake of abbreviation in many places, where the intention should be clear or the distinction is not crucial. 
\begin{notation}
    For every agent $i$ and allocation $x$, 
    \[\sigma_x^{\uparrow(\downarrow)}(i):= \{j \in [m] | i \in s^{\uparrow(\downarrow)}_j(x)\}\]
\end{notation}

Next, we want to define the partial derivative of $\sum_{i=1}^n f(\ov{i}{x})$ w.r.t. alternative $j \in [m]$.
\begin{definition}[marginal contributions]
    Let $x \in \Delta^m$. For every $j \in [m]$,
    \[mc^{\uparrow(\downarrow)}_j(x):=\sum_{i \in s^{\uparrow(\downarrow)}_j(x)}f'(\ov{i}{x})\]
\end{definition}
Note that $mc_j(x)$ depends on $x_j$ only through $s_j$, however the contribution of each member to the sum is a function of their overall satisfaction $\ov{i}{x}: = \sum_j \min(x^i_j,x_j)$. In case $x^i_k=x_k$ for some $k \in [m]$ (whether $k=j$ or other), $f'(\ov{i}{x})$ does not exist, and we take the right or left derivative accordingly. 
\begin{proposition}[MRS condition]\label{lemma_mrs}
$x \in \arg \max_{y \in \Delta^m}\sum_i f(\ov{i}{y})$ if and only if
\[ mc^\uparrow_j(x) \leq mc^\downarrow_k(x)\quad \forall j,k \in [m]\]    
\end{proposition}
Note that when $s^\uparrow_j=s^\downarrow_j$ and $s^\uparrow_k=s^\downarrow_k$, marginal contributions also coincide and the two inequalities for $j$ and $k$ are reduced to the equation $mc_j=mc_k$. Intuitively, optimal allocations must admit the MRS demand because if $mc^\uparrow_j(x) > mc^\downarrow_k(x)$ for some $j$ and $k$, increasing $x_j$ at the expense of $x_k$ would increase $\sum_i f(\ov{i}{})$, up to the point where the MRS condition is satisfied. (And, every local maximum of a concave function is also a global one).

\subsection{Axiomatic Properties of CTRs}
We start with examining the satisfaction of the axiomatic demands mention above for CTRs. It is easy to observe that \textbf{EFF} and \textbf{PAR} are satisfied by all members in the class. As for other demands, their performance is not as satisfying.  
\begin{proposition}
    No CTR satisfies \textbf{SP}. 
\end{proposition}
 \begin{proof}
     Consider the following counter example. Let $n=m=2$, $x^1=(.5,.5)$ and $x^2=(0,1)$. Let $x=\tr_f(\vx)$. Since $\tr_f$ satisfies \textbf{RR}, we know that $x=(y,1-y)$ for some $y \in [0,0.5]$, so that $s_1=1$, $s_2=2$, $\ov{1}{x}=y +0.5$ and $\ov{2}{x}=1-y$. Thus, 
     \[f'(y+0.5)=mc_1=mc_2=f'(1-y) \implies y=.25\]
     ($f'$ is strictly decreasing, thus injective). Now assume agent $1$ misreports $\hat{x}^1=(1,0)$ instead, and call the new outcome $(z,1-z)$. Then 
     \[f'(z)=mc_1=mc_2=f'(1-z) \implies z=.5\]
     and agent $1$'s satisfaction has increased from $0.75$ to $1$. 
 \end{proof}

When it comes to proportional fairness, the only CTR that achieves any such demand (for $m >2$) is the logarithmic rule where $f=\ln$, aka the Nash Product Rule, that satisfies the relatively strong \textbf{AFS}. That stands in alignment with the discrete model where the PAV rule that maximizes $\sum_i \mathcal{H} \Big(|A_i \cap W|\Big)$ (where $\mathcal{H}$ is the harmonic function) is the only Thiele rule that admits a somewhat weaker demand called EJR \cite{aziz2017justified}. The proof of Nash Rule's \textbf{AFS} property is dismissed here as a particular case of Theorem \ref{thm:lambda_AFS}. On the contrary,  we show that it is the only CTR that satisfies the much weaker \textbf{PROP}.
\begin{proposition}
    For $m > 2$, the Nash rule $\ \tr_{\ln}$ is the only CTR that satisfies \textbf{PROP}.
\end{proposition}
\begin{proof}
    Let $\vx$ be a single-minded preferences vector, and $T_f(\vx)=x$. By MRS, $s_jf'(x_j)=s_kf'(x_k) \forall j,k$ (note that under single-minded profiles $mc^\uparrow_j=mc^\downarrow_j$ for all $j$ such that $0 < x_j <1 $). If $x$ satisfies \textbf{PROP} then $x_j=\frac{s_j}{n}$, yielding 
    \[s_jf'(\frac{s_j}{n})=s_kf'(\frac{s_k}{n}) \implies \frac{s_j}{n}f'(\frac{s_j}{n})=\frac{s_k}{n}f'(\frac{s_k}{n})\]
    Satisfying this for general $n,\ 1 \leq s_j,s_k \leq n$ means that $xf'(x)$ is a constant function, thus $f=\ln$.
    \end{proof}
Moreover, the logarithmic CTR does not satisfy \textbf{CS}.
\begin{example}
    Let $m=3$ and $[n]$ consisting of 3 disjoint sets $n_1,n_2,n_3$ such that $n_1=n_2=0.3n,\ n_3=0.4n$. Each of the subsets is homogeneous with preferences $x^1=(1,0,0),x^2=(.5,.5,0),x^3=(0,0,1)$ respectively.
Then $\tr_{\ln}(\vx)=x=(.5,0, .5)$. However, with $y=(0.55,0.05)$, $|y|=\frac{n_1+n_2}{n}$, every agent $i \in n_1 \cup n_2$ has $\ov{i}{y}=0.55 > \ov{i}{x}=0.5$.
\end{example}
 Interestingly, the Nash product does output a core solution under preference model not too far from $\ell_1$ \cite{aziz2019fair}. Whether the core is always non-empty, and what rule can find core solutions when they exist, remains an open question.

\section{Utilitarian vs. Egalitarian Welfare}
\rmr{maybe say something about being unimpressed with the axiomatic angle for CTRs, and looking for a more explicit tradeoff. } 
After reviewing the not-very-impressive axiomatic performance of CTR rules, we develop in this section a more explicit presentation of the compromise one has to make between utilitarianism and distributive fairness. We start with the notion of \textit{Inequality Aversion} captured in a concave function $f$.  \rmr{use subsections here e.g. for utilitarian / egalitarian / proportionality results or upper/lower bounds.}
\begin{definition}[ Inequality Aversion]
    For any twice differentiable $f: [0,1] \to \R$, the \textit{Inequality Aversion} of $f$, $IAV_f: (0,1] \to \R$ is 
    \[IAV_f(t)=-\frac{tf'(t)}{f''(t)}\]
\end{definition}
More commonly, $-\frac{tf'(t)}{f''(t)}$ is known as the "Relative Risk Aversion" of $f$, a major concept in decision making under uncertainty \cite{dyer1982relative}. While the two contexts are unrelated, the technical similarity is comprehensible. In principle, the concavity of $f$ implies $\frac{1}{n}\sum_i f(\ov{i}{x}) \leq f\Big(\frac{1}{n}\sum_i \ov{i}{x}\Big)$, and $-\frac{tf'(t)}{f''(t)}$ is a proxy for how big the gap is. The higher $-\frac{tf'(t)}{f''(t)}$ is, the less dispersed will be the optimal distribution for $\{\ov{i}{x}\}_{i \in [n]}$. In decision making under uncertainty, $f$ represents utility and $\frac{1}{n}\sum_i f(\ov{i}{x})$ is the expectancy $E\big[f(L)\big]$ for some random variable $L$ distributed over $\{\ov{i}{x}\}_{i \in [n]}$, whereas in our case $f$ is a function \textit{chosen } by a social planner to promote her normative balance between fairness (low disparity) and overall welfare $\sum_i f(\ov{i}{x})$. Here are some examples for the IAV of different functions.

\begin{center}
\begin{tabular}{ c|c|c|c } 
 & $f$ & $IAV_f$ &\\ 
 \hline
 & $-t^{-p},\ \scriptstyle{p>0} $ & $ = 1+p\ \forall t \in (0,1]$ &\\
 & $ -e^{t^{-p}},\ \scriptstyle{p>0}$ & $ \geq 1+p\ \forall t \in (0,1] $ & \\
   & $ t^p,\ \scriptstyle{1>p>0}$ & $ = 1-p\ \forall t \in (0,1] $ &\\
   & $\ln(t) $ & $ =1\ \forall t \in (0,1] \hspace{0.55cm} $ & \\
   & $t(2-t) $ & $ \leq \frac{1}{4}\ \forall t \in (0,1] \hspace{0.55cm} $ &
\end{tabular}
\end{center}

The lemma below shows an equivalent representation for $IAV$ that will later come handy. Then, Example \ref{ex:IAV_disjnt_groups} that follows will provides a clear intuition to its major role here. When writing $IAV = (\leq) (\geq) \lambda$, we mean that the corresponding relation holds in all $[0,1]$.

\begin{lemma}
The following are equivalent for any twice differentiable concave function $f: [0,1] \to \mathbb{R}$: 
\begin{itemize}
    \item $IAV_f \stackrel{(\leq)}{\geq} \lambda$ for some $\lambda > 0$. 
    \item $\forall \alpha > 1$, $\frac{f'(t)}{f'(\alpha t)} \stackrel{(\leq)}{\geq} \alpha^{\lambda}$
\end{itemize}
\end{lemma}
\begin{proof}
     Essentially, $-\frac{tf''(t)}{f'(t)} \stackrel{(\leq)}{\geq} \lambda$ means that $f'(t)$ decreases faster (slower) than $t^{-\lambda}$:
    \[\dv[]{}{t}\Big[t^\lambda f'(t)\Big]=t^{\lambda-1}\big(\lambda f'(t)+tf''(t)\big) \leq 0 \iff -\frac{tf''(t)}{f'(t)} \geq \lambda\] 
     And if $t^\lambda f'(t)$ is non-increasing (non-decreasing), then 
    \[\forall \alpha > 1,\ \ t^\lambda f'(t) \stackrel{(\leq)}{\geq} (\alpha t)^\lambda f'(\alpha t) \iff \frac{f'(t)}{f'(\alpha t)}\stackrel{(\leq)}{\geq}  \alpha^{\lambda} \]
\end{proof}

\begin{example}\label{ex:IAV_disjnt_groups}
    Consider a single-minded profile for $m=2$ where $s_2 > s_1$. MRS at the outcome $x=\tr_f(\vx)$ gives
\begin{gather*}
    mc_1  = s_1f'(\ov{1}{x})=s_1f'(x_1) \\
         mc_2  = s_2f'(\ov{2}{x})=s_2f'(x_2) \\
         \implies \frac{f'(x_1)}{f'(x_2)} = \frac{s_2}{s_1}
\end{gather*}   
   
     If $\lambda$ bounds $IAV_f$, it also bounds the extent to which $\tr_f$ favors the majority $s_2$. If $IAV_f \geq \lambda$, 
    \[\big(\frac{x_2}{x_1}\big)^\lambda \leq \frac{f'(x_1)}{f'(x_2)}  \implies \frac{\ov{2}{x}}{\ov{1}{x}}= \frac{x_2}{x_1} \leq \big(\frac{s_2}{s_1}\big)^{1/\lambda} \]
    In particular, for $\lambda \to \infty$ we get the egalitarian maxmin allocation $x_1=x_2$.  
    For $IAV_f  \leq \lambda$ we will have the inequalities reversed :
    \[\big(\frac{x_2}{x_1}\big)^\lambda \geq \frac{f'(x_1)}{f'(x_2)}  \implies \frac{\ov{2}{x}}{\ov{1}{x}}= \frac{x_2}{x_1} \geq \big(\frac{s_2}{s_1}\big)^{1/\lambda}\]
    so that at the limit $\lambda \to 0$ the welfare maximizing allocation $x=(1,0)$ is achieved. 
    Finally, if $IAV=1$ (meaning $f=\ln$) the budget is allocated proportionally, $x_j=\frac{s_j}{s_j+s_k}$. 
\end{example}
   Our main results, presented in this section, build on the idea demonstrated in \textbf{Example \ref{ex:IAV_disjnt_groups}} to provide fairness $/$ welfare guarantees within the CTR class.
   
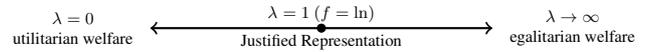
\begin{figure}\label{fig:CTR_line}
       \centering
\begin{tikzpicture}[scale=0.75, every node/.style={scale=0.7}]
\draw[thick,<->] (-3,0) -- (3,0);
\filldraw[black] (0,0) circle (2pt);
\node[align=center,anchor=south] at (0,0) { $\lambda=1\ (f=\ln)$};
\node[align=center,anchor=north] at (0,0) { \small Justified Representation};
\node[align=center,anchor=west] at (3.2,0) { $\lambda \to \infty$\\
\small egalitarian welfare};
\node[align=center,anchor=east] at (-3.2,0) { $\lambda=0$\\
\small utilitarian welfare};
\end{tikzpicture}
       \label{fig:enter-label}
   \end{figure}

   We can figuratively map this class to $(0,\infty)$ based on the $IAV$ parameter, as demonstrated in Figure \ref{fig:CTR_line}. At the left limit $\lambda =0$ we have the welfare maximizing rule $\ar(\vx)=\arg \max \sum_i \ov{i}{}$. Theorem \ref{thm:wf_bound} bounds the welfare loss as a function of the upper bound on the $IAV$. On the other hand, for $1 \geq IAV \geq \lambda$ we get increasingly close approximations of \textbf{AFS} as coming closer to the Nash Product Rule where $IAV=1$ (Theorem \ref{thm:lambda_AFS}), and at the right limit $IAV \geq \lambda \to \infty$ we have the egalitarian rule (Theorem \ref{thm:egl_bound}). In other words $0 < IAV < 1$ correspond to rules that favor utilitarian over egalitarian welfare (to various degrees), $IAV > 1$ means favoring egalitarianism, and the Nash rule with $IAV=1$ is point of "exact balance" between the two.
   
\subsection{Welfare Loss}
Theorem \ref{thm:wf_bound} below shows the welfare preserved when the $IAV$ is small enough, and is preceded by some necessary 
definitions. 
 \begin{definition}
    The \textbf{Welfare} in allocation $x$ is  $W(x):=\sum_i \ov{i}{x}$. 
\end{definition}
\begin{definition}
    The \textbf{Welfare loss} of allocation $x$ is defined 
    \[\wl(x)=1-\frac{W(x)}{\max_{y \Delta^m} W(y)}\]    
\end{definition}
\begin{notation}
    Given two allocations $x$ and $y$, we write:
    \begin{align*}
        J_x:= \{j \in [m]| x_j \geq y_j\}\ &;\ J_y:= \{j \in [m]| x_j < y_j\}\\
        \delta_j:= |x_j-y_j|\ \forall j \in [m]\ &; \text{ and } \delta:=\sum_{j \in J_x} \delta_j=\sum_{j \in J_y} \delta_j
    \end{align*}  
    \end{notation}
  The difference in satisfaction between $x$ to $y$ for every agent is thus bounded by 
  \[\ov{i}{y}-\ov{i}{x} \leq \sum_{k \in J_y\cap\sigma_x^\uparrow(i)}\delta_k - \sum_{j \in J_x\cap\sigma_x^\downarrow(i)}\delta_j\]
  Indeed, an agent might gain less then all of $\delta_k$ for $\ k \in J_y\cap\sigma_x^\uparrow(i)$ if $x_k < x^i_k < y_k$, and may lose some of $\delta_j$ even if $j \notin J_x\cap\sigma_x^\downarrow(i)$ in case $y_j < x^i_j < x_j$. 
  
  \begin{theorem}\label{thm:wf_bound}
    Let $x=\tr_f(\vx)$ such that $IAV_f \leq \lambda$. Then 
    \[\wl(x) \leq  \frac{\lambda m^\lambda}{\lambda m^\lambda+\lambda +1}\]
\end{theorem}
 \begin{proof}
     Let $x = \tr_f(\vx)$ and $y = \arg \max\sum_i \ov{i}{}$. Then      
     \begin{align*}     
       W(y)-W(x) \leq  & \sum_{k \in J_y} s^\uparrow_k(x)\delta_k - \sum_{j \in J_x} s^\downarrow_j(x)\delta_j\\
         \leq &\ s^\uparrow_{\max}\sum_{k \in J_y}\delta_k - \sum_{j \in J_x} s^\downarrow_j\delta_j\\
         = & \sum_{j \in J_x}(s^\uparrow_{\max}-s^\downarrow_j(x))\delta_j\\
        \leq & \sum_{j \in J_x}(s^\uparrow_{\max}-s^\downarrow_j(x))x_j
     \end{align*}
     where $s^\uparrow_{\max}:= \max_{k \in J_y}s^\uparrow_k$ and we used $\sum_{k \in J_y} \delta_k=\sum_{j \in J_x}  \delta_j$ and $\delta_j \leq x_j$.
     Now, by MRS and $IAV_f \leq \lambda$ we have for all $j,k \in [m]$: 
     \begin{gather*}
          s^\uparrow_k f'(1) \leq mc^\uparrow_k \leq mc^\downarrow_j \leq s^\downarrow_j f'(x_j)\\
          \implies s^\uparrow_k \leq s^\downarrow_j \frac{f'(x_j)}{f'(1)} \leq s^\downarrow_j \big(\frac{1}{x_j}\big)^{\lambda}
     \end{gather*}
     and therefore $\sum_{j \in J_x}(s^\uparrow_{\max}-s^\downarrow_j(x))x_j \leq s^\uparrow_{\max}\sum_{j \in J_x}(1-x_j^{\lambda})x_j$.
     The function $g(t)=(1-t^\lambda)t$ is concave and has a maximum $\frac{\lambda}{(\lambda+1)^{\frac{\lambda+1}{\lambda}}}$ at $t^*=(\lambda+1)^{-\frac{1}{\lambda}}$. Thus, 
    \[\sum_{j \in J_x}(1-x_j^{\lambda})x_j \leq \frac{\sum_{j \in J_x}x_j}{t^*}g(t^*)=\sum_{j \in J_x}x_j \frac{\lambda}{\lambda +1} \leq \frac{\lambda}{\lambda +1} \]
  Now,  
     \begin{align*}
          W(x) & = \sum_i \ov{i}{x} \geq \sum_i \sum_{j \in \sigma(i)}x_j = \sum_j s^\downarrow_j x_j\\
         & \geq s^\uparrow_{\max}\sum_j x_j^\lambda \cdot x_j 
          \geq s^\uparrow_{\max} m\Big(\frac{1}{m}\Big)^{1+\lambda}\\
          & = s^\uparrow_{\max} \Big(\frac{1}{m}\Big)^{\lambda}
     \end{align*}     
     Therefore,
     \begin{align*}
         \frac{W(y)-W(x)}{W(x)} & \leq m^\lambda\frac{\lambda}{\lambda +1}\\
         \implies \frac{W(y)-W(x)}{W(y)} & \leq \frac{m^\lambda\frac{\lambda}{\lambda +1}}{1+m^\lambda\frac{\lambda}{\lambda +1}}= \frac{\lambda m^\lambda}{\lambda m^\lambda+\lambda +1}
     \end{align*}
     
 \end{proof}

\begin{corollary}[Welfare loss in single minded profiles]
   In the special case of single-minded preferences (where every voter allocates the full budget
   to some unique alternative $j \in [m]$), the utilitarian allocation puts all budget on the alternative with highest support, meaning $W(y)=s^\uparrow_{\max}$, yielding \rmr{equality or inequality?}
   \[\wl(x)=\leq \frac{s^\uparrow_{\max}\sum_{j \in J_x}x_j\frac{\lambda}{\lambda +1}}{s^\uparrow_{\max}} \]
   Moreover, since $s_jf'(x_j)=s_kf'(x_k)$ means $x_k \geq x_j \iff s_k \geq s_j$, if $s_k=s^\uparrow_{\max}$ then $x_k =\max_j x_j \geq \frac{1}{m}$ and thus $\sum_{j \in J_x}x_j =\sum_{j \neq k}x_j \leq \frac{m-1}{m}$. Hence, 
   \[\wl(x) \leq \frac{m-1}{m}\frac{\lambda}{\lambda +1}\]
\end{corollary}
\rmr{to we have any lower bound on WL at all?}
\subsection{Approximate AFS}
 \jwr{
\begin{lemma}\label{lemma:min_agent}
    Let $x=\tr_f(\vx)$ such that $IAV_f \geq \lambda$. Then 
    \[\min_i \ov{i}{x} \geq \frac{1}{m}\Big(\frac{1}{n}\Big)^{\frac{1}{\lambda}}\]
\end{lemma}
Not that when $\lambda \to \infty$ we get $\min_i \ov{i}{x} \geq \frac{1}{m}$ that is the individual share guarantee of the egalitarian rule (assuming $n >m$).

\begin{theorem}[$\lambda$-AFS]
    Let $x=\tr_f(\vx)$ such that $\lambda_1 \leq IAV_f \leq \lambda_2$ for some $\lambda_1 \leq 1 \leq \lambda_2$. Then,
    For every $\alpha$-cohesive set $s \in [n]$
    \[ \frac{1}{|s|}\sum_{i \in s}\ov{i}{x} \geq  \alpha^{\frac{1}{\lambda}}\]
\end{theorem}
\begin{proof}
  Let an $\alpha$-cohesive group $s \subset [n]$. Since $|\bigcap_{i \in s}x^i| \geq \alpha \geq \alpha^{\frac{1}{\lambda}}$, if $\bigcap_{i \in s}x^i$ if covered by $x$ we are done. Otherwise, there exists $j \in [m]$ such that $\min_{i \in s}x^i_j > x_j$ and, since $f'$ is convex, 
\[mc^\uparrow_j = \sum_{i \in s^\uparrow_j}f'(\ov{i}{x}) \geq \sum_{i \in s}f'(\ov{i}{x}) \geq sf'(\ov{s}{})\] 
where $\ov{s}{}:= \sum_{i \in s}\ov{i}{x}$
By MRS, $sf'(\ov{s}{}) \leq mc^\downarrow_k\ \forall k \in [m]$. Thus, we would now like to bound $\min_{k}mc^\downarrow_k$. To do that, Let $\sigma(i)=\{k \in [m]| i \in s^\downarrow_k\}$ for every agent $i$ and apply Lemma \ref{lemma:min_agent} to get
\begin{align*}
    \sum_k mc_k \cdot x_k & = \sum_k x_k \sum_{i \in s_k} f'(\ov{i}{x}) = \sum_i \sum_{k \in \sigma(i)}x_kf'(\ov{i}{}) \\
    & \leq \sum_i \sum_{k \in \sigma(i)}x_kf'\Big(\frac{1}{mn^{1/\lambda_1}}\Big) \leq nf'\Big(\frac{1}{mn^{1/\lambda_1}}\Big) \\
    & \leq n \big(mn^{1/\lambda_1}\big)^{\lambda_2}f'(1)=m^{\lambda_2}n^{\frac{\lambda_2+\lambda_1}{\lambda_1}}f'(1)
\end{align*}
 Therefore, as $\sum_k x_k =1$, there must exist $k \in [m]$ such that $mc_k \leq m^{\lambda_2}n^{\frac{\lambda_2+\lambda_1}{\lambda_1}}f'(1)$. Hence, 
\begin{gather*}
   sf'(\ov{s}{}) \leq m^{\lambda_2}n^{\frac{\lambda_2+\lambda_1}{\lambda_1}}f'(1) \implies \frac{s}{n} \leq \frac{f'(1)}{f'(\ov{s}{})} \leq (\ov{s}{})^\lambda
\end{gather*}
Overall, we showed $\ov{s}{} \geq \min \left(\alpha, \big(\frac{s}{n}\big)^{\frac{1}{\lambda}}\right) \geq \alpha^{\frac{1}{\lambda}}$.
\end{proof}

 }

The next Theorem shows how good an approximation of AFS we can guarantee when increasing the (lower bound on) $IAF$ towards $1$. 
\begin{theorem}[$\lambda$-AFS]\label{thm:lambda_AFS}
    Let $x=\tr_f(\vx)$ such that $1 \geq IAV_f \geq \lambda$. Then,
    For every $\alpha$-cohesive set $s \in [n]$
    \[ \frac{1}{|s|}\sum_{i \in s}\ov{i}{x} \geq  \alpha^{\frac{1}{\lambda}}\]
\end{theorem}
\begin{proof}
  Let an $\alpha$-cohesive group $s \subset [n]$. Since $|\bigcap_{i \in s}x^i| \geq \alpha \geq \alpha^{\frac{1}{\lambda}}$, if $\bigcap_{i \in s}x^i$ if covered by $x$ we are done. Otherwise, there exists $j \in [m]$ such that $\min_{i \in s}x^i_j > x_j$ and, since $f'$ is convex, 
\[mc^\uparrow_j = \sum_{i \in s^\uparrow_j}f'(\ov{i}{x}) \geq \sum_{i \in s}f'(\ov{i}{x}) \geq sf'(\ov{s}{})\] 
where $\ov{s}{}:= \sum_{i \in s}\ov{i}{x}$
By MRS, $sf'(\ov{s}{}) \leq mc^\downarrow_k\ \forall k \in [m]$. Thus, would now like to bound $\min_{k}mc^\downarrow_k$. To do that, Let $\sigma(i)=\{k \in [m]| i \in s^\downarrow_k\}$ for every agent $i$ and note that $\ov{i}{x} \geq \sum_{k \in \sigma(i)}x_k$. Now,
\begin{align*}
    \sum_k mc_k \cdot x_k & = \sum_k x_k \sum_{i \in s_k} f'(\ov{i}{x}) = \sum_i \sum_{k \in \sigma(i)}x_kf'(\ov{i}{}) \\
    & \leq \sum_i \sum_{k \in \sigma(i)}x_kf'\left(\sum_{k \in \sigma(i)}x_k\right) \leq n f'(1)\\
\end{align*}
where the last inequality is due to the fact that $tf'(t)$ is an increasing function by $IAV_f \leq 1$. As $\sum_k x_k =1$, there must exist $k \in [m]$ such that $mc_k \leq nf'(1)$. Hence, 
\begin{gather*}
   sf'(\ov{s}{}) \leq nf'(1) \implies \frac{s}{n} \leq \frac{f'(1)}{f'(\ov{s}{})} \leq (\ov{s}{})^\lambda
\end{gather*}
Overall, we showed $\ov{s}{} \geq \min \left(\alpha, \big(\frac{s}{n}\big)^{\frac{1}{\lambda}}\right) \geq \alpha^{\frac{1}{\lambda}}$.
\end{proof}

\subsection{Egalitarian Loss and Individual Shares}
We move on now to discuss the egalitarian guarantees we can provide for high enough $IAV$. We  aim on arguing that $\tr_f$ approaches the egalitarian rule at the limit $IAV_f  \to \infty$. First, we have the next result on individual share. 
\begin{proposition}\label{propos:ind_shares}
     Let $x=\tr_f(\vx)$ such that $IAV_f \geq \lambda$ and $\ov{\min}{x}:=\min_i \ov{i}{x}$. Then 
     \[\ov{\min}{x} \geq \frac{1}{1+(m-1)(n-1)^\frac{1}{\lambda}}\]
\end{proposition}
\begin{proof}
    Since every agent must be in $s^\uparrow_j$ for at least one $j \in [m]$ (unless her satisfaction is $1$), there exist $j \in [m]$ such that $s^\uparrow_j \geq f'(\ov{\min}{x})$. Thus, due to the MRS conditions,
    \begin{gather*}
        \forall k \in [m],\quad f'(\ov{\min}{x}) \leq mc^\downarrow_k \leq (n-1)f'(\min_{i \in s_k}\ov{i}{x})\\
        \implies (n-1)^\frac{1}{\lambda}\ov{\min}{x} \geq \min_{i \in s_k}\ov{i}{x} \geq x_k
    \end{gather*}
    and in particular, $\ov{\min}{x} \geq x_j$. Thus, 
    \[1 = x_j +\sum_{k \neq j}x_k \leq \ov{\min}{x}\Big(1+(m-1)(n-1)^\frac{1}{\lambda}\Big)\]
\end{proof}
For $\lambda \to \infty$ that bound is $1/m$ which is $\max_{y \in \Delta^m}\min_i\ov{i}{y}$ \underline{in the worst case}. The following definition formalizes "being close to egalitarian rule" by its full meaning. 
\begin{definition}
    The \textbf{\emph{egalitarian loss}} of allocation $x$ is defined 
    \[\el(x):=1-\frac{\min_i\ov{i}{x}}{\max_{y \in \Delta^m}\min_i\ov{i}{y}}\] .
\end{definition}
Nevertheless, Proposition \ref{propos:ind_shares} alone does give an upper bound on the egalitarian loss for single-minded profiles. 
\begin{corollary}[egalitarian loss in single-minded profiles]
    The egalitarian allocation for single minded profiles is uniform (assuming all projects $j$ have non-empty support $s_j$), making $\ov{i}{}=1/m\ \forall i$. Thus, by Proposition \ref{propos:ind_shares} 
    \[\el(x) = 1 - \frac{\ov{\min}{x}}{1/m} \leq 1- \frac{m}{1+(m-1)(n-1)^\frac{1}{\lambda}}\]
\end{corollary}
Bounding $\el(x)$ for general profiles, is, however, much more challenging. Our way of doing that can be roughly presented as follows.  The egalitarian rule allocation is characterized by the fact the no improvement on \textit{all} argmin agents is possible, in other words a shift in any direction will harm (or will not benefit) some of them. The proof of Theorem \ref{thm:egl_bound} below will show that for large enough $IAV$, a unanimous improvement is not feasible on all agents with satisfaction \textit{close to minimal}, thereby limiting the increase in minimum satisfaction. Formalizing that requires the following notion.    
\begin{definition}[Directional Derivative]
    For every agent $i$ an two allocations $x,y \in \Delta^m$, define the derivative of $\ov{i}{}$ towards $y$ at $x$ as 
    \[\pdv{\ov{i}{x}}{(y-x)} := \dv[]{}{\alpha}\ov{i}{}(\alpha y +(1-\alpha)x) \Big |_{\alpha=0}\]
\end{definition}
Note that
\begin{align*}
    \pdv{\ov{i}{x}}{(y-x)} & = \dv[]{}{\alpha}\Big[\sum_j \min(\alpha y_j +(1-\alpha)x_j,x^i_j)\Big]\\
    & = \sum_{k \in J_y\cap\sigma_x^\uparrow(i)}\delta_k - \sum_{j \in J_x\cap\sigma_x^\downarrow(i)}\delta_j\\
    & \geq \ov{i}{y}-\ov{i}{x}
\end{align*}

We are now ready to this part's main result.
\begin{definition}
    For all $m,n,\lambda,$ let
    \[\gamma(m,n,\lambda):= \max_{\omega \in [0,1]}\min\left(m\omega\ , 1- \left(\frac{\omega}{n-1}\right)^{\frac{1}{\lambda}}\right)\]
\end{definition}
 \begin{theorem}\label{thm:egl_bound}
Let $x = \tr_f(\vx)$ where $IAV_f \geq \lambda$. Then 
 \[\el(x) \leq \gamma(m,n,\lambda)\]
 \end{theorem}
 Note that $\gamma(m,n,\lambda) \leq 1$ and that $\lim_{\lambda \to \infty}\gamma(m,n,\lambda)=0$ for all $m,n$. Table \ref{tab:gamma_values} shows some values of $\gamma(m,n,\lambda)$.
 
\begin{table}
    \centering
    \begin{tabular}{lllll} 
\toprule
$m$ & $\lambda=0.1$ & $\lambda=1$ & $\lambda=10$ & $\lambda=100$ \\
\midrule
3 & 1.000 & 0.997 & 0.474 & 0.079 \\
8 & 1.000 & 0.999 & 0.519 & 0.087 \\
12 & 1.000 & 0.999 & 0.537 & 0.090 \\
20 & 1.000 & 0.999 & 0.558 & 0.094 \\
\bottomrule
\end{tabular}
    \caption{Different values of $\gamma(m,n,\lambda)$ for $n=100$.}
    \label{tab:gamma_values}
\end{table}
 
 \begin{proof}
      and $y$ the maxmin allocation. Since $x$ is an optimum of $\sum_i f(\ov{i}{})$, 
     \[0 \geq \pdv{}{(y-x)}\left[\sum_i f(\ov{i}{})\right]=\sum_i f'(\ov{i}{})\pdv{\ov{i}{x}}{(y-x)}\]
     Now let us add some notations. Let $s_1$ be all agents $i$ for which $\ov{i}{y} > \ov{i}{x}$, and $s_2:=[n] \setminus s_1$. Denote $\rho:= \min_{i \in s_2}\ov{i}{x}$ and $\mu:=\min_{i \in \arg \min_i \ov{i}{x}}\pdv{\ov{i}{x}}{(y-x)}$. Surely $\arg \min_i \ov{i}{x} \subseteq s_1$, therefore 
     \begin{gather*}
         0 \geq \pdv{}{(y-x)}\left[\sum_i f(\ov{i}{})\right] \geq f'(\ov{\min}{x}) \cdot \mu -(n-1)f'(\rho)\\
         \implies \frac{n-1}{\mu} \geq \frac{f'(\ov{\min}{x})}{f'(\rho)} \geq \left(\frac{\rho}{\ov{\min}{x}}\right)^\lambda
     \end{gather*}
     Since $\rho=\ov{i}{x}$ for some agent in $s_2$, $\ov{\min}{y} \leq \rho$. And, $\ov{\min}{y}-\ov{\min}{x} \leq \mu$ because the directional derivative is greater then the actual gain in satisfaction for every agent. Thus, 
     \[(n-1)\left(\frac{\ov{\min}{x}}{\ov{\min}{y}}\right)^\lambda \geq \ov{\min}{y}-\ov{\min}{x}\]
Thus, if $\ov{\min}{y}-\ov{\min}{x}= \omega \in [0,1]$, then $\el(x) = \frac{\omega}{\ov{\min}{y}} \leq m\omega$ because $\ov{\min}{y} \geq \frac{1}{m}$, and on the other hand, $\el(x) = 1- \frac{\ov{\min}{x}}{\ov{\min}{y}} \leq 1- \left(\frac{\omega}{n-1}\right)^{\frac{1}{\lambda}}$. Thus, 
\[\el(x) \leq \min\left(m\omega\ , 1- \left(\frac{\omega}{n-1}\right)^{\frac{1}{\lambda}}\right) \leq \gamma(m,n,\lambda)\]
 \end{proof}


 \section{Discussion}
\begin{figure}\label{fig:graph}

\pgfplotsset{compat=1.18}

\begin{tikzpicture}
    \begin{semilogxaxis}[
        xlabel=\(\lambda\),
        ylabel=Loss,
        domain=1/100:100,
        samples=500,
        grid=both,
        minor grid style={gray!25},
        major grid style={gray!50},
        width=8cm,
        height=4cm,
       legend pos=north west,
        title = {(a) Loss bounds on general profiles}
    ]
        \addplot[green, thick, name path=WLh] {x*15^x/(x*15^x+x+1)};
        \addplot[green, thick, dashed, name path=WLl] {x*3^x/(x*3^x+x+1)};
          \addplot[purple, thick, smooth] coordinates {
        (0.01,1) ( 0.1 , 1.0 )
( 0.5 , 0.9999995410681068 )
( 1 , 0.9993225548781104 )
( 2 , 0.9743636414550525 )
( 3 , 0.9148777586032315 )
( 4 , 0.8450582191804875 )
( 5 , 0.7791178951537928 )
( 6 , 0.7196675156927942 )
( 7 , 0.6669011612100486 )
( 8 , 0.6211161655581108 )
( 9 , 0.5814352428652492 )
( 10 , 0.5457445943510613 )
( 20 , 0.34184776921784166 )
( 30 , 0.25095437004893395 )
( 40 , 0.19874051453157116 )
( 50 , 0.16516516516516516 )
( 60 , 0.14214611591272608 )
( 70 , 0.1244679954313771 )
( 80 , 0.11110878038130689 )
( 90 , 0.10073488051438484 )
( 100 , 0.09113584898104488 )
( 200 , 0.04932013344631958 )
    };
\addplot[purple, thick,dashed,  smooth] coordinates {

( 0.01 , 1.0 )
( 0.1 , 1.0 )
( 0.5 , 0.9996922129886119 )
( 1 , 0.9827195616669301 )
( 2 , 0.8759685327931293 )
( 3 , 0.7625867923090903 )
( 4 , 0.6704037724294394 )
( 5 , 0.597707272804981 )
( 6 , 0.5399282474324897 )
( 7 , 0.4924924924924925 )
( 8 , 0.45345345345345345 )
( 9 , 0.42042042042042044 )
( 10 , 0.39201170172620314 )
( 20 , 0.23925598399944836 )
( 30 , 0.17506873433491665 )
( 40 , 0.13931453415317374 )
( 50 , 0.11640162122561193 )
( 60 , 0.10004803312412602 )
( 70 , 0.08802509467155539 )
( 80 , 0.07867457018476487 )
( 90 , 0.07146388158038497 )
( 100 , 0.06536716162475908 )
( 200 , 0.036036036036036036 )

};

        
    \end{semilogxaxis}
\end{tikzpicture}
\begin{tikzpicture}
    \begin{semilogxaxis}[
        xlabel=\(\lambda\),
        ylabel=Loss,
        domain=1/100:100,
        samples=500,
        grid=both,
        minor grid style={gray!25},
        major grid style={gray!50},
        width=8cm,
        height=4cm,
       legend pos=north west,
        title = {(b) Loss bounds on single-minded profiles}
    ]
       
        \addplot[green,thick, name path=WLhs] {(14/15))*x/(1+x)};
        \addplot[green, thick, dashed, name path=WLls] {(2/3))*x/(1+x)};
          \addplot[purple,  thick,name path=ELl]  {1-15/(1+14*(100-1)^(1/x))};
          \addplot[purple,  thick,dashed, name path=ELl]  {1-3/(1+2*(20-1)^(1/x))};
        
    \end{semilogxaxis}
\end{tikzpicture}
    \caption{Green lines show welfare loss upper bounds,  scarlet lines show Egalitarian loss. Thick lines are  for $m=15$ and $n=100$, dashed lines for $m=3$ and $n=20$.  }
\end{figure}
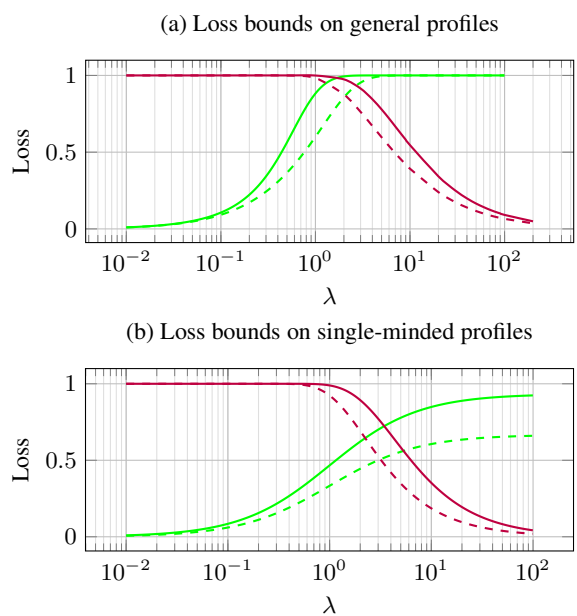

 This article presents a less common approach to the well known welfare-fairness trade-offs introduced in almost every Social Choice problem. Instead of separately testing the performance of different rules \cite{elkind2023settling,aziz2019fair}, we offered a class of rules that almost none of them properly satisfy common axioms, but rather introduce a continuum of quantifiable trade-off points between the contradicting desires. From a practical point of view, such presentation of the range of possibilities might be more appealing and convenient to work with for social planners. Graph (a) in Figure \ref{fig:graph} compares our results for welfare and egalitarian
loss as functions of $\lambda$.

 As this is a work in progress, we surely hope for possible improvements in some of our results. For example, we presented significantly lower bounds for welfare and egalitarian losses under single-minded profiles (Figure \ref{fig:graph} graph (b)). Intuitively, however, in single-minded profiles the interests of different agents are as contradictory as they could be, leading to the conjecture that they might actually be the worst possible cases. We thus hope for some room for improvement in the results for general profiles as well. Moreover, some potentially interesting directions still await for further investigation. E.g., we did not look for approximate strategy-proofness in this article. Intuitively, what makes the strategy-proofness of the utilitarian rule is that $mc_j=s_j$ for all $j$, making it impossible for an agent $i \in s_j$ to force an increase in $x_j$ towards $x^i_j$ via manipulating her vote. On the other hand, the $IAV$ of a function determines the impact that low satisfaction has compared to size in $mc_j=\sum_j f'(\ov{i}{})$. Thus, we might expect a bound on the agent's ability to increase $mc_j$ that depends on the $IAV$.

\bibliography{mybibfile}

\end{document}